\documentclass[onecolumn,11pt]{article}
\usepackage[top=1in, bottom=1in, left=1in, right=1in]{geometry}
\usepackage{epstopdf, epsfig}
\usepackage{amssymb,amsfonts,amsmath}
\usepackage{graphicx}
\usepackage{cancel}
\usepackage[switch]{lineno}
\usepackage{lettrine}
\usepackage{soul,xcolor}
\setstcolor{red}
\usepackage{xfrac}
\usepackage{bm}
\usepackage{color}
\usepackage{float}
\usepackage[bottom,flushmargin,hang,multiple]{footmisc}
\usepackage{natbib}
\usepackage{setspace}
\newcommand\blfootnote[1]{%
  \begingroup
  \renewcommand\thefootnote{}\footnote{#1}%
  \addtocounter{footnote}{-1}%
  \endgroup
}

\definecolor{mPurp}{RGB}{126 , 47, 142}
\definecolor{mRed}{RGB}{217 , 83, 25}
\definecolor{mBlue}{RGB}{0 , 114, 189}
\definecolor{mGreen}{RGB}{119 , 172, 48}

\newcommand{\up}[0]{\widetilde{\bm{u}}}
\newcommand{\uv}[0]{\bm{u}'}
\newcommand{\um}[0]{\overline{\bm{u}}}
\usepackage{lettrine}

\newcommand{\fref}[1]{{Fig.~\ref{fig:#1}}}

\newcommand{\fl}[1]{{{\small \sffamily{(#1)}}}}

\title{\LARGE{\textbf{Near-Wake Dynamics of a Vertical-Axis Turbine}}}

\author{\normalsize{Benjamin Strom$^{*}$, Brian Polagye, Steven L. Brunton}\\
\footnotesize{Department of Mechanical Engineering, University of Washington,
Seattle, WA 98195, USA}
}
\date{}
\begin{document}
\maketitle
\blfootnote{$^*$ Corresponding author (ben@xflowenergy.com).}

\vspace{-.15in}
\begin{abstract}
Cross-flow, or vertical-axis, turbines are a promising technology for capturing kinetic energy in wind or flowing water and their inherently unsteady fluid mechanics present unique opportunities for control optimization of individual rotors or arrays. 
To explore the potential for beneficial interactions between turbines in an array, coherent structures in the wake of a single two-bladed cross-flow turbine are examined using planar stereo particle image velocimetry in a water channel. 
First, the mean wake structure of this high chord-to-radius ratio rotor is described, compared to previous studies, and a simple explanation for observed wake deflection is presented. 
Second, the unsteady flow is then analyzed via the triple decomposition, with the periodic component extracted using a combination of traditional techniques and a novel implementation of the optimized dynamic mode decomposition. The latter method is shown to outperform conditional averaging and Fourier methods, as well as uncover frequencies suggesting a transition to bluff-body shedding in the far wake. 
Third, vorticity and finite-time Lyapunov exponents are then employed to further analyze the oscillatory wake component. 
Vortex streets on both sides of the wake are identified, and their formation mechanisms and effects on the mean flow are discussed. 
Strong axial (vertical) flow is observed in vortical structures shed on the retreating side of the rotor where the blades travel downstream. 
Time-resolved tracking of these vortices is performed, which demonstrates that vortex trajectories have significant rotation-to-rotation variation within one diameter downstream. This variability suggests it would be challenging to harness or avoid such structures at greater downstream distances.\\

\noindent\textbf{Key words:}  Cross-flow turbine, coordinated array control, triple decomposition, dynamic mode decomposition, finite-time Lyapunov exponents
\end{abstract}

\section{Introduction}
\label{sec:intro}
Cross-flow, or vertical-axis, turbines are experiencing a resurgence in research interest for the conversion of wind and water currents to electricity. 
One motivation is the mounting evidence that arrays of closely-spaced cross-flow turbines can extract more energy per unit land area than industry-standard axial flow turbines. 
This property has benefits where the array mounting area is limited, such as roof-top installations, and where the region of high flow speed is concentrated, such as mountain passes or tidal channels. 
\cite{dabiri_2014} and~\cite{dabiri_2015} reported a power density of 10-20 W/m$^2$, compared to the 1-3 W/m$^2$ output of conventional axial-flow wind turbine farms~\citep{adams_2013, mackay_2008}. 
Notably, in field experiments, \cite{brownstein_2016} find average rotor performance in an array was 20\% higher than the performance of a single isolated turbine. 
Similarly,~\cite{strom_2018,scherl_2021} have demonstrated a 30\% increase in average rotor output in an array of two cross-flow turbines compared to isolated turbine performance. 
The remarkable performance of densely packed cross-flow turbines stems from several fluid mechanical phenomena. 
First, the orientation of the rotation axis results in an acceleration of the bypass flow, especially on the side of the rotor where the blades are retreating (traveling downstream). 
Neighboring rotors placed in this flow benefit from the increased incident mean velocity. 
Second, the tip vortices shed from the blades have an axis of rotation that lie in a plane parallel to the ground. 
These vortices induce vertical mixing, increasing the transfer of momentum from the high-speed flow above the array to the rotor level, increasing the stream-wise wake recovery rate~\citep{bachant_2015b}. 
Finally, we speculate that performance may be enhanced through the interaction between periodic coherent structures shed by an upstream turbine and the blades of a downstream turbine. 
This hypothesis is inspired by schooling fish who have been shown to benefit from well-timed interactions with vortices shed from upstream individuals~\citep{whittlesey_2010,maertens_2017}. 
These potential performance increase mechanisms motivate the study of the mean and periodic components of a cross-flow turbine wake, with a special focus on coherent structures that may interact with nearby turbines in an array and be exploitable through control.

Measurement and analysis of cross-flow turbine wakes have been been conducted for decades, starting with~\cite{muraca_1976}. 
Point measurements using Pitot tubes~\citep{muraca_1976, battisti_2011}, hot wires~\citep{battisti_2011, bergeles_1991, peng_2016, persico_2016}, laser~\citep{buchner_2018}, acoustic~\citep{bachant_2015b, kinzel_2012, kinzel_2015}, and Doppler velocimetry have been used to describe the mean wake structure, spectra, and time-average turbulence statistics. 
Two-component~\citep{araya_2015, araya_2017, eboibi_2016, posa_2016} and three-component~\citep{hohman_2018, rolin_2015, tescione_2014} planar particle image velocimetry (PIV) and magnetic resonance velocimetry~\citep{ryan_2016} measurements, as well as simulations~\citep{boudreau_2017, nini_2014, scheurich_2011, scheurich_2013, shamsoddin_2014}, have been used to investigate the wake spatial variability, including wake geometry, recovery rate, and the role of turbulence. 
Despite the widely varying rotor configurations and operating conditions across these studies, a set of features common to cross-flow turbine wakes have emerged.  
First, wake measurements have often been made at mid-plane of the rotor, perpendicular to the rotation axis. 
In this plane, all studies report some asymmetry or angular deflection of the wake in the direction of turbine rotation, with a more intense shear layer on side where the blades are advancing (traveling upstream. 
Flow structures shed at the blade passing frequency (rotational frequency $x$ number of blades) have been identified in nearly all studies on the retreating side of the wake (e.g.,~\cite{brochier_1986,bachant_2015b, battisti_2011, ryan_2016}) and on both the retreating and advancing sides (e.g.,~\citep{posa_2016, araya_2017, boudreau_2017, hohman_2018}). 
\cite{araya_2017} determined that these shear flow oscillations transition to those corresponding to a bluff body in the far wake. 
Second, areas of high turbulence intensity have been identified in a streak on the advancing side of the wake deficit~\citep{bachant_2015b} and on both sides~\citep{hohman_2018, rolin_2015}. 
Third, studies that examined the three-dimensional wake structures have identified the primary mechanism for wake recovery as axial (vertical) flow induced by vortices shed from the blade tips~\citep{kinzel_2012, kinzel_2015, boudreau_2017} or the induced cross-stream (horizontal) flow~\citep{bachant_2015b}. 
In contrast, the wake recovery in axial-flow turbines is driven primarily by turbulent mixing upon the breakdown of the helical tip vortices~\citep{lignarolo_2015, boudreau_2017}. 
Consequently, wake recovery rates have been documented to be significantly faster than those of axial-flow turbines~\citep{boudreau_2017, dabiri_2011}. 
In addition to wake measurements, multiple studies have performed measurements within the rotor, demonstrating the importance of dynamic stall and subsequent blade-vortex interactions in normal cross-flow turbine operation~\citep{brochier_1986,fujisawa_2001,ferreira_2009, edwards_2015, eboibi_2016,dave_2021}.

The contributions of the present study are three-fold. 
First, time-average, three-component, planar PIV measurements are presented for a relatively high chord-to-radius ratio turbine and the wake structure is related to the turbine rotor hydrodynamics. 
Second, we demonstrate that an algorithm incorporating the dynamic mode decomposition (DMD) can identify energetically important modes that cannot be discovered by other methods. 
Third, by analyzing the form and trajectory of coherent structures shed into the near-wake, we identify the region over which their propagation is deterministic, which is of relevance to array control. 

\section{Cross-Flow Turbine Operation and Experimental Methods}
\subsection{Cross-Flow Turbine Background}

\begin{figure}
\centering
\includegraphics[width=\textwidth]{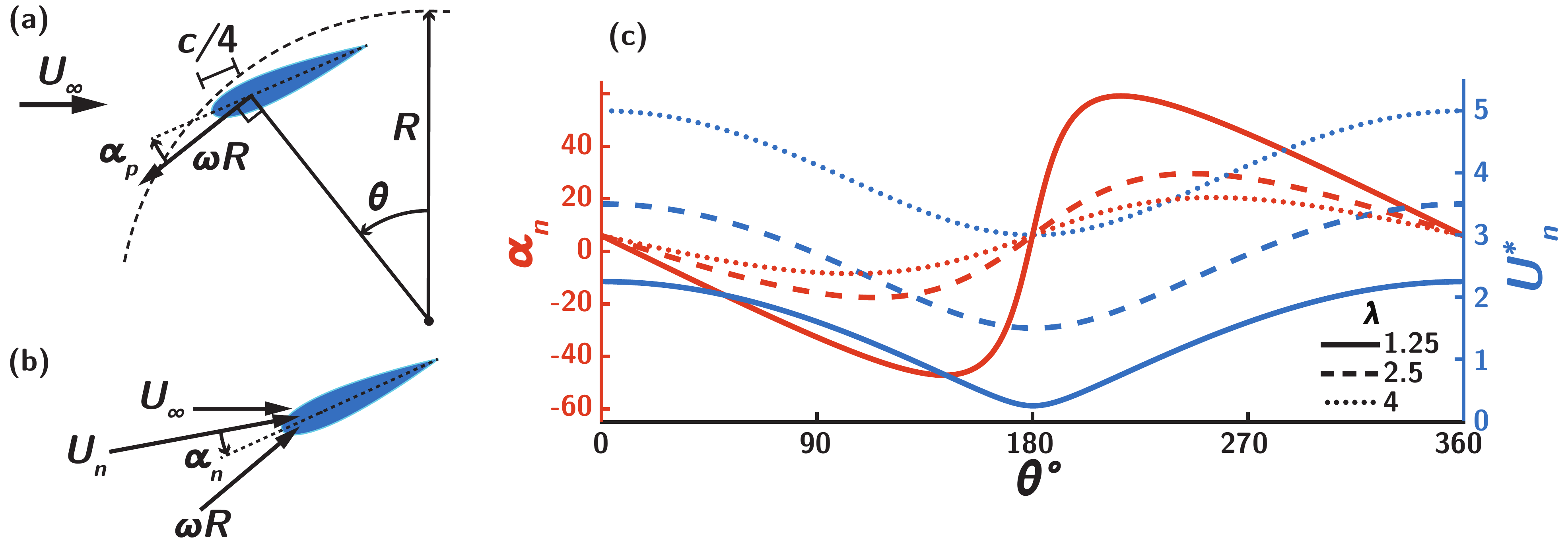}
\caption{(a) Diagram of geometric and kinematic quantities (b) Free stream, rotational, and resulting total velocity vector and local angle of attack. (c) Variation in local angle of attack and flow velocity as a function of azimuthal blade position for three values of tip-speed ratio ($\lambda$).}\label{fig:intro_dia}
\end{figure}

Despite typically having only a single degree of freedom, rotation about a central axis, the fluid dynamics of cross-flow turbines are inherently unsteady. 
This is because, even with a steady inflow, the local flow conditions experienced by the blade vary cyclically over the course of a single rotation. 
Neglecting flow variations induced by the turbine, the local flow velocity magnitude and angle of attack vary according to
\begin{equation}\label{eq:U_n}
	U_n(\theta)^* = \frac{| \bm{U_n}(\theta) |}{U_\infty} = \sqrt{\lambda^2+2\lambda \cos(\theta)+1} 	
\end{equation}
and
\begin{equation}\label{eq:alpha_n}
		\alpha_n(\theta) = -\text{Tan}^{-1}\left[\sin(\theta), \lambda + \cos(\theta)\right]+\alpha_p,
\end{equation}
respectively, where $\text{Tan}^{-1}$ is the four quadrant arctangent and $\alpha_p$ is the preset pitch (blade mounting) angle, $\theta$ is the blade azimuthal position, and $\lambda$ is the tip-speed ratio, or non-dimensional rotation rate. $\lambda$ is given by
\begin{equation}
	\lambda = \frac{\omega R}{U_\infty},\label{eq:tsr}
\end{equation} 
where $\omega$ is the angular velocity of the turbine, $R$ is the radius, and $U_\infty$ is the free stream velocity. 
As shown in Fig.~\ref{fig:intro_dia}, we define $\theta = 0$ where the quarter-chord of the blade is traveling directly upstream. 
Vector diagrams of these quantities, as well as examples of how they vary over the course of one rotation are given in Fig.~\ref{fig:intro_dia}. 
Especially at low tip-speed ratios, the local angle of attack can far exceed the static stall angle of the foil. 
These kinematics, equivalent to a rapid pitching maneuver, can lead to dynamic stall and the corresponding roll-up of a leading edge vortex~\citep{Eldredge2019arfm}. 
Depending on the timing, strength, and trajectory, this vortex may contribute to or detract from power output~\citep{strom_2015, ferreira_2009}. 
Dynamic stall and the resulting coherent structures provide an opportunity for optimization of blade-fluid structure interactions, either in the case of a single turbine, as in~\cite{strom_2017}, or for multi-rotor interactions. 
Array optimization that seeks to maximize power based on the mean flow has been successfully demonstrated~\citep{brownstein_2016}. 
However, arrays of cross-flow turbines may also be able to take advantage of periodic coherent structures in the wakes of nearby turbines~\citep{strom_2018,scherl_2021}. 
In addition to their potential influence on array interactions, the role of coherent  wake structures in deficit recovery rate motivates the close examination of their lifetime and trajectory.

The efficiency with which a cross-flow turbine converts flow kinetic energy to rotational mechanical energy is given by
\begin{align}
	C_P = \frac{\omega \tau}{\frac{1}{2} \rho U_\infty^3 A},
\end{align}
where $\omega$ is the turbine rotation rate, $\tau$ is the mechanical torque produced by the rotor, $\rho$ is the operating fluid density, and $A$ is the rotor swept area. 
Performance is often characterized as a function of the tip-speed ratio in Eq.~\eqref{eq:tsr} and can be presented as a time-average or phase-averaged quantity~\citep{polagye2019}. 
For wake measurements to be most meaningful, it is useful to operate the turbine at a realistic operating condition, such as the tip-speed ratio that yields the maximum $C_P$, because wake characteristics differ significantly between power producing and non-power producing operating states~\citep{araya_2015}.

At large turbine scales, such as those used for commercial power production, rotor geometries with few, relatively small chord-length blades exhibit high maximum efficiency. 
One example is the Sandia 34 m test bed turbine, with a peak $C_P$ of approximately 0.41~\citep{ashwill_1992}. 
However, because cross-flow turbine performance can improve rapidly with increasing Reynolds number~\citep{bachant_2016b, miller_2018}, at smaller scales it can be useful to increase the chord-length of the turbine to maximize the blade Reynolds number. 
Larger chord-length foils may also be more structurally robust, which is important because blade fatigue is often the cause of cross-flow turbine structural failure~\citep{mollerstrom_2019}. 
Finally, large chord-to-radius turbines have a lower tip-speed ratio at peak performance, reducing losses from support structures and radiated noise. 
These factors motivate the study of cross-flow turbines with relatively high chord-to-radius ratios. 
The low tip-speed ratio at peak efficiency of this geometry results in large local angle-of-attack variations (Fig.~\ref{fig:intro_dia}) and in separation and stall when operating at maximum $C_P$~\citep{snortland2019}. 

\subsection{Flume and Turbine}
Experiments were performed in the Alice C. Tyler flume at the University of Washington. 
The flume has a test section measuring 0.76 m wide and 4.9 m long. 
The dynamic water depth was 0.47 m, the free stream velocity ($U_{\infty}$) was maintained at 0.7 m/s, and the turbulence intensity ($\text{rms}(\bm{u}')/U_\infty$) was 1.5\%. 
The water temperature was held at a constant 16.3$\pm$0.4$^\circ$C.

\begin{figure}
\centering
\includegraphics[width=0.425\textwidth]{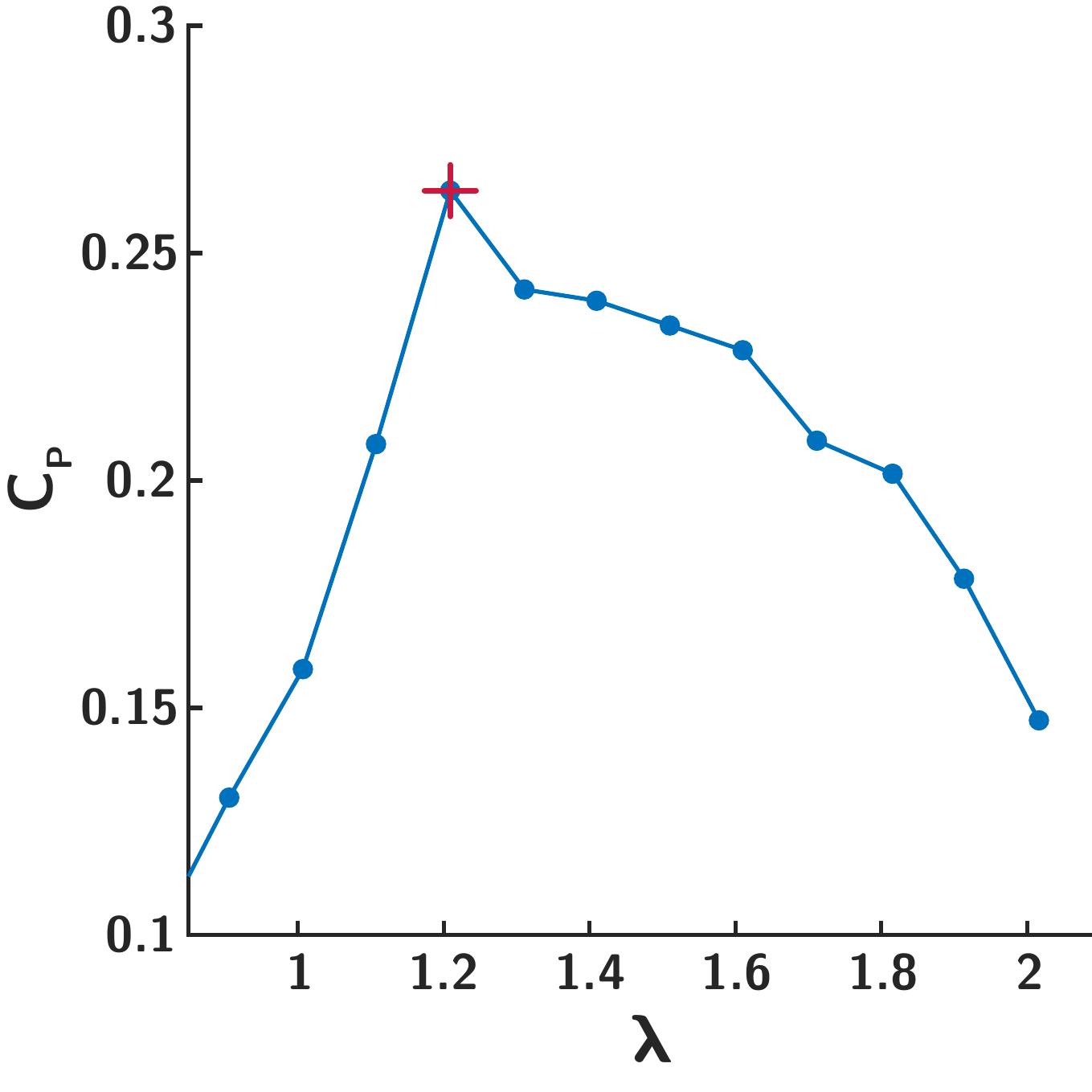}
\vspace{-.1in}
\caption{Performance curve (mechanical efficiency versus tip-speed ratio) for the experimental turbine. Red cross indicates operating point during wake data collection. Performance curve data was collected using the experimental setup detailed in~\cite{strom_2017}, but during the PIV experiments with a cantilevered turbine (Fig.~\ref{fig:setup}), the upper load cells was removed to increase the stiffness of the experimental setup.}\label{fig:perf_curve}
\vspace{-.1in}
\end{figure}

The cross-flow turbine model had a height of $H =  0.234$ m and a diameter of $D = 0.172$ m. 
The diameter-based Reynolds number,
\begin{equation}
Re_D = \frac{D U_\infty}{\nu},
\end{equation}
 was $1.1\times 10^5$, where $\nu$ is the water kinematic viscosity. 
 The turbine was vertically centered in the flume with a depth-based Froude number
 \begin{equation}
 Fr = \frac{U_\infty}{\sqrt{g d}},
 \end{equation}
 was 0.33 where $g$ is the acceleration due to gravity and $d$ is the dynamic water depth. 
 The blockage ratio, or the ratio of the turbine cross-sectional area to the test section cross-sectional area, was 11\%. 
 The turbine consisted of two, straight, NACA0018 profile blades with chord length of $c=0.061$ m for a chord-to-radius ratio of 0.71 and solidity, 
 \begin{equation}
 	\sigma = \frac{N c}{\pi D},
\end{equation}
of 0.225, where $N$ is the blade count. 
The blades were mounted to a 0.012 m diameter central shaft via circular end-plates at a pitch angle of 6 degrees (leading edge rotated outwards about the quarter-chord). 
The turbine was operated under constant angular velocity control at its peak performance point of $C_P = 0.26$ at a tip-speed ratio of $\lambda = 1.2$. 
The turbine performance curve is given in Fig.~\ref{fig:perf_curve}, and details on the methods used to determine turbine performance can be found in~\cite{strom_2017}.

The turbine was cantilevered from the face of a direct-mount servomotor (Yaskawa SGMCS) with an integrated 1,048,576 edges-per-revolution encoder providing blade position feedback, which was recorded via a counter on a National Instruments PCIe data acquisition card to a computer at a rate of 1 kHz. 
This acquisition was synchronized with the PIV measurement system. 
The servomotor regulated the rotational speed of the turbine to a constant value and power generated was actualized as reverse current in the servomotor and dissipated in a dump resistor. 
The turbine rotor and servomotor were mounted to a robotic gantry system, providing accurate translation of the rotor in the stream-wise direction. 
\subsection{PIV Measurement}

\begin{figure}
\centering
\includegraphics[width=\textwidth]{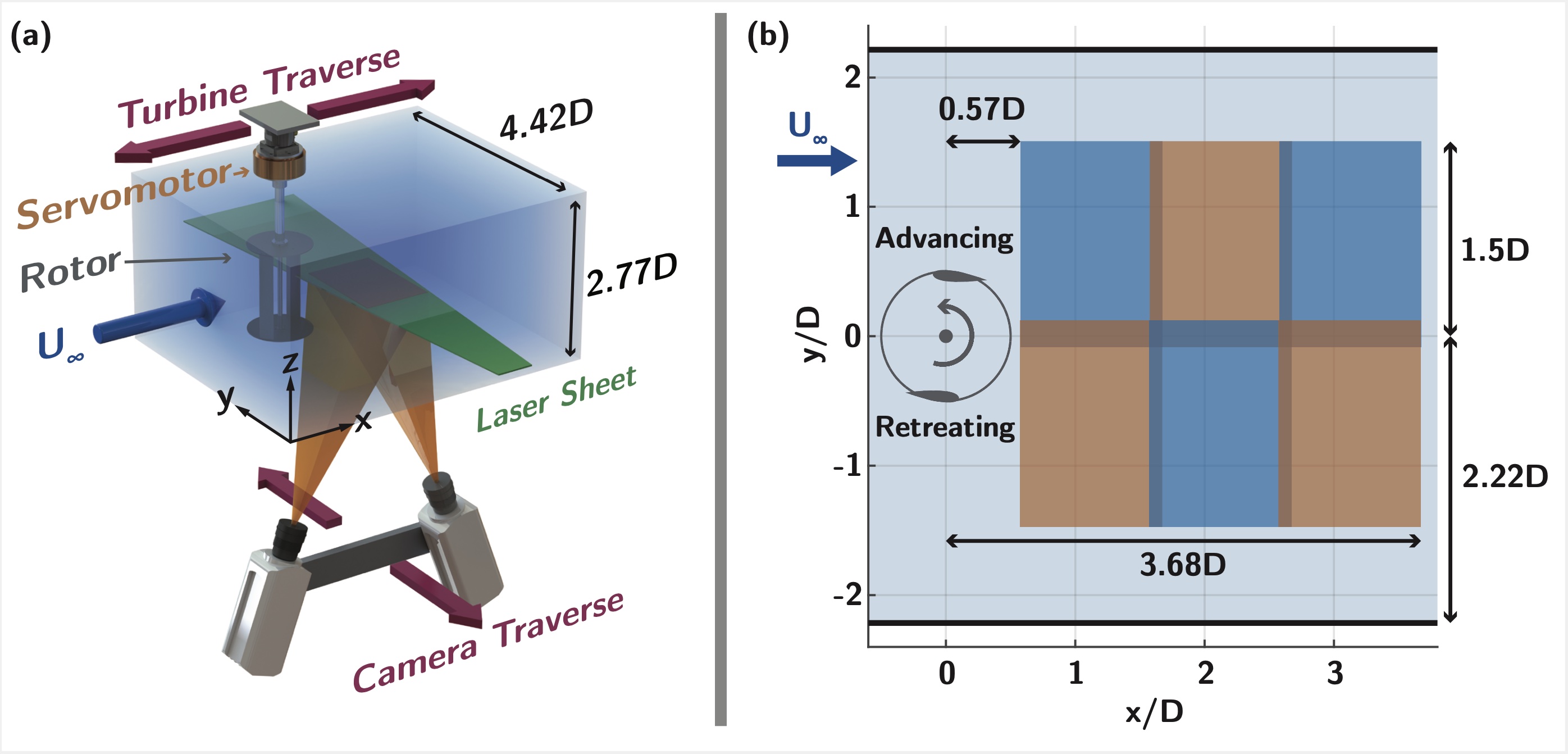}
\caption{Turbine and PIV measurement setup diagram (a) and PIV measurement locations in the mid-plane along the $\bm{z}$ direction (b).}\label{fig:setup}
\end{figure}

Measurements of the turbine wake were acquired using time-resolved stereo planar PIV. 
Data was collected in a free-running manner at 100 Hz, corresponding to 5.35 degrees of blade rotation between measurements, and was not locked to specific blade positions. 
Measurements were taken at the mid-span of the turbine, in the plane normal to the axis of rotation. 
Illumination was provided by a Continuum TerraPIV Nd:YLF laser, and images were captured by two Phantom V641 cameras, with resolutions of $2560 \times 1600$ pixels. 
Cavitation bubbles from the flume recirculation pump were used as passive tracers and measured approximately 1.5 pixels in diameter. 
Measurement resolution was increased by capturing the wake using six overlapping fields-of-view, as illustrated in Fig.~\ref{fig:setup} (a). 
We note that data is collected in these six regions in separate experiments, and therefore not synchronized.  
Consequently, we align the data in post-processing using the algorithm discussed in \textsection \ref{sec:periodic_structures} and presented in more detail in ~\cite{Nair2020prf}. 
The combined measurement area, shown in Fig.~\ref{fig:setup} (b), started 0.57$D$ downstream from the turbine axis, and extended 3.68$D$ downstream, and 3$D$ in the cross-stream direction. 

Spatial calibration was performed with custom stereo calibration target spanning the entire width of the flume section in conjunction with a robotic camera gantry used to repeatably move the cameras in the cross-stream direction. 
Post-processing was performed with custom image manipulation software and TSI Insight for the cross-correlation. 
Ghost velocities due to small laser-sheet / calibration target misalignment was corrected through image warping in post processing. 
Velocity fields were calculated using iterative multi-grid processing, with initial square interrogation window side size of 64 pixels and a final size of 16 pixels. 
With 50\% window overlap, the resulting velocity vector spacing was 0.0068$D$.

\section{Results and Analysis: Mean Flow}

The mean wake deficit contours and normalized velocities are shown in \fref{mean_wake}. 
As in prior work, we observe an asymmetric wake deficit with an intense shear layer on the advancing side of the wake (see Fig.~
\ref{fig:setup} \fl{b} for advancing versus retreating nomenclature). 
Wake deficit recovery occurs faster on the retreating side, as previously observed by~\cite{tescione_2014}. 
The mean wake deficit is never negative, meaning there is no recirculation region. 
\cite{araya_2017} showed a decrease in wake deficit with reduction in the number of blades; however, even their two-bladed turbine showed some negative streamwise velocity. 
A survey of wake wake measurements in prior work indicates that neither rotor efficiency, solidity, nor the expression of dynamic solidity of~\cite{araya_2017} are good predictors of whether or not a negative wake deficit occurs. 
It is possible that some combination of these factors, in addition to the test section blockage ratio and rotor thrust, would be necessary to predict the magnitude of the wake deficit. 

\begin{figure}
\centering
\includegraphics[width=\textwidth]{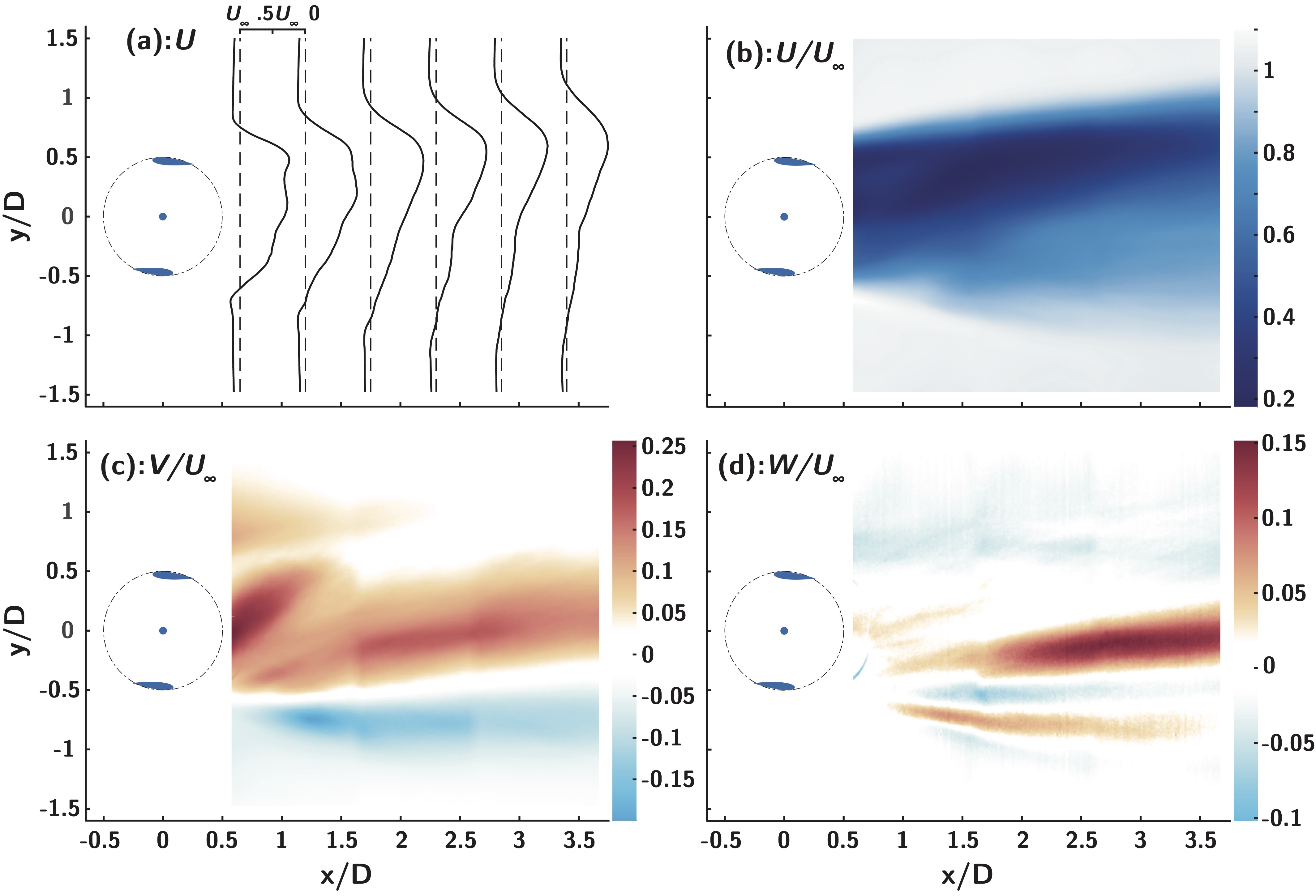}
\vspace{-.35in}
\caption{\fl{a} Mean wake deficit profiles. Streamwise velocity profiles along cross-stream stations (dashed lines). The distance from one station to the next is a change in velocity equivalent to the mean freestream velocity, $U_\infty$. \fl{b} The mean streamwise, \fl{c} cross-stream, and \fl{d} vertical (axial) velocities, normalized by the freestream velocity.}\label{fig:mean_wake}
\end{figure}

Despite the differences in turbine geometry, the streamwise wake velocity is similar to those described by~\cite{peng_2016} (five blades, $c/R = 0.3$) and~\cite{hohman_2018} (three blades, $c/R = 0.2$), suggesting that rotor geometry has limited effect on the mean wake structure. 
The blockage ratio of 11\% also increases the shear between the wake and bypass flow compared to an unconfined case, but, as noted in~\cite{ross2020experimental}, the time-average wake structure is qualitatively invariant with this magnitude of blockage.

There are several conflicting theories in the literature about the root cause of the mean wake profile asymmetry. Specifically:
\begin{itemize}
\item \cite{araya_2017}: ``In all cases, there is a notable asymmetry of the [cross-flow turbine] wake. This is attributed to the stronger shear layer that forms on the side of the turbine where the blades are advancing upstream."
\item \cite{hohman_2018}: ``This behavior is as expected, as the majority of the power is generated on the advancing side of the turbine, and therefore a larger momentum deficit will be seen on this side."
\item \cite{bachant_2015b}: compares the effect to that of a rotating cylinder, stating ``Compared with the rotating cylinder wake measurements of \cite{lam_2016} we see a similar asymmetry in the mean streamwise velocity. 
The wake is less asymmetrical with respect to the wake centreline for the turbine compared to the rotating cylinder for the same nondimensional rotation rate, although some of these differences may be due to the cylinder experiments' lower Reynolds numbers.''
\item \cite{peng_2016}: provide multiple explanations: ``There are two major factors that may contribute to this wake asymmetry. One factor is that more turbulent structures are produced at the windward than at the leeward. When the blade advances under adverse pressure gradients at the windward, stronger vortex shedding and much severer flow separations take place. The other factor is that the wake flows are transported toward the windward. First, when the blade moves upwind at the windward, it causes stronger blockage effect compared to that at the leeward. Therefore, at the windward, the blade wake is characterized by a lower pressure, which induces the cross-wind flows. Second, when the blade operates at the downstream half-revolution, the strong angular momentum drags and propels the wake flows toward the windward."
\end{itemize}

\begin{figure}
\centering
\includegraphics[width=\textwidth]{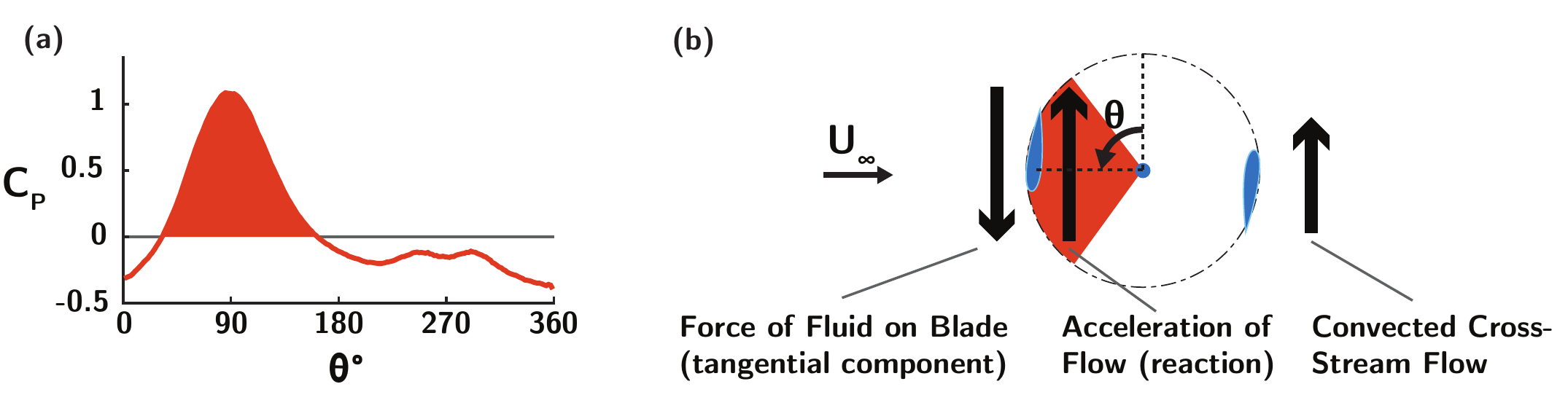}
\vspace{-.35in}
\caption{(a) Measured power coefficient as a function of azimuthal blade position for a single blade in a cross-flow turbine. (b) Diagram of the force exerted on the blade by the fluid necessary for power production, the resulting reaction force and acceleration of flow, and the resulting convected cross-stream flow which leads to wake deflection.}\label{fig:wake_deflect}
\end{figure}
 
We propose a simpler explanation for wake asymmetry. Power measurements on a single-bladed cross-flow turbine~\citep{strom_2017} indicate that the majority, it not all of the power is produced on the upstream side of the rotor, with peak power production centered at approximately $\theta = 90^\circ$ (immediately eliminating the explanation of~\cite{hohman_2018}). 
This is illustrated in \fref{wake_deflect} \fl{a}. 
Torque production arises from the force acting on the blade in the tangential direction, which in the region of peak power, corresponds to the $-\widehat{y}$ direction in the coordinate system used here. This results in a time-average rotor force in this direction~\citep{polagye2019}. 
As a consequence of this application of force, the fluid must experience a force in the opposite direction, specifically, in the $+\widehat{y}$ direction. 
This effect is analogous to the angular velocity induced during axial-flow turbine operation, where the induced flow is opposite the direction of turbine rotation~\citep{burton_2001}. 
In the case of the cross-flow turbine, the flow velocity induced in the cross-stream direction is advected downstream through the rotor to the wake, as depicted in \fref{wake_deflect} \fl{b}. 
Strong evidence of this cross-stream velocity is seen in \fref{mean_wake} \fl{c}, though the action of blade tip vortices could also induce flow in this direction~\citep{battisti_2011}. 

Returning to the properties of the mean wake, it is curious to note significant vertical (axial) velocities present in \fref{mean_wake} \fl{d}. 
Because we are sampling on the mid-plane and the turbine rotor is symmetric about this plane, one would expect the wake to reflect this vertical symmetry, resulting in no out-of-plane velocities. 
However,~\cite{peng_2016} and~\cite{rolin_2015} both observe similar asymmetries. 
Interactions with the free-surface or the flume floor boundary layer could be mechanisms responsible for this phenomenon, though the former is unlikely as mid-plane vertical flows have been observed in wind-tunnel measurements.  
The stability of coherent wake structures may play some role in this asymmetry, as described later. 

\section{Results and Analysis: Periodic Structures}
\label{sec:periodic_structures}

Flows with natural or forced periodicity, such as the wake of a cross-flow turbine, contain fluctuations that are regular in space or time, and thus should not be characterized as turbulence. 
It is then useful to analyze flows with periodic, organized content in terms of the triple decomposition of~\cite{hussain_1970}:
\begin{equation}
\bm{u}(\bm{x},t) = \um(\bm{x}) + \up(\bm{x},\phi(t)) + \uv(\bm{x},t),
\end{equation}
where the total flow, $\bm{u}$, is the superposition of a time-averaged flow, $\um$, the periodic flow, $\up$, parameterized by the phase $\phi(t)$, and turbulent fluctuations, $\uv$. 
While $\um$ is calculated through simple time averaging, there are multiple approaches for separating the periodic and turbulent components. 
For flows where the forcing mechanism or flow periodicity are measured simultaneously with the velocity field, $\up$ can be calculated by phase-averaging (i.e. by computing the ensemble mean of measurements occurring at the same phase of the forcing oscillator). 
This was the approach taken by the pioneers of the triple decomposition~\citep{hussain_1970}, and is commonly employed, for example in the wake of an axial-flow wind turbine by~\cite{eriksen_2017}. 
Drawbacks to this approach include the necessity of measuring the phase of the forcing oscillator, as well as the introduction of statistical uncertainty in the case that measurements are not locked to the forcing oscillator~\citep{cantwell_1983}. 
This is the case for our measurements because PIV data collection was free-running and not locked to the turbine blade position, but was time-synchronized with measurements of turbine performance and blade position. 
A potential solution for the uncertainty in free-running measurements is to use a weighted average based on the phase offset for a given measurement from the phase in question. 

Alternatively, Fourier averaging, where $\up$ is estimated using a truncated Fourier series~\citep{sonnenberger_2000}, eliminates the need to measure the forcing signal simultaneously with flow measurements and removes error associated with phase uncertainty. 
However, since the base forcing frequency must be known or assumed, periodic flow structures due to phenomena other than the primary forcing mechanism may not be included in $\up$. 

The desire to automatically extract and rank the importance of spatially coherent and temporally periodic flow phenomena at multiple scales, without a priori knowledge of the frequencies of interest, has inspired a number of methods. 
Instead of an oscillatory component composed of a single base frequency, these methods yield a triple decomposition of the form
\begin{equation}\label{eq:mm_triple}
\bm{u}(\bm{x},t) = \um(\bm{x}) + \sum\limits_{n=1}^R \up_n(\bm{x},\phi_n(t)) + \uv(\bm{x},t),
\end{equation}
where $R$ is the number of oscillatory modes used in the reconstruction and $n$ is the mode number. 
Because the frequency, amplitude, and phase of oscillations are determined directly from velocity time series, this potentially reduces errors inherent to conditional averaging of free-running data and DFT methods.

The dynamic mode decomposition (DMD)~\citep{schmid2010jfm,rowley_2009,tu2014jcd,kutz_2016} provides a scalable and data-driven approach to extract the periodic component $up_n$ in \eqref{eq:mm_triple}.  
DMD is a combination of proper orthogonal decomposition (POD)~\citep{berkooz_1993,holmes_2012,taira2017aiaa,Taira2020aiaaj} in space and the Fourier transform in time.  
Although POD is widely used for spatial mode extraction~\citep{edgington_2014, oberleithner_2011}, including in the wake of an axial-flow turbine~\citep{premaratne_2016, lignarolo_2015}, the modes are known to mix frequency content. 
This is illustrated by the fact that the snapshot POD modes~\citep{sirovich:1987,aubry1988dynamics,Brunton2019book} do not depend on the order of the flow data time-series.  
Many instances of the failure of POD to extract dynamically important modes for multi-scale systems have been documented~\citep{sayadi_2012,baj_2015,taira2017aiaa}.  
In contrast, DMD modes are a linear combination of POD modes, specifically designed to be coherent in space and have distinct oscillation frequencies, as well as growth or decay rates. 

\subsection{Approaches for Triple Decomposition}
Here, we describe several triple decomposition methods.  
In the next section, we will compare the efficacy of these methods for detecting oscillatory structures and their performance based on error and energy capture. 

\subsubsection{Blade Position Conditional Averages}

We compute three conditional averages based on the turbine blade position at the time of PIV image capture. 
In these methods, PIV data is binned based on blade position. 
Subsequently, the median, mean, or weighted mean flow field velocities are calculated. 

For a single point in space, all $n$ measurements are collected for which the blade position, $\theta$, satisfies
\begin{equation}
\left| \theta - \theta_i \right| < \Delta \theta,
\end{equation}
where $\Delta \theta$ is the half bin width and $\theta_i$ denotes the $i^{\mathrm{th}}$ bin center.  
The bin mean is
\begin{equation}
\up(\bm{x},\theta_i) = \frac{1}{n} \sum\limits_{j = 1}^n \left(\bm{u}(\bm{x},\theta_j)-\um(\bm{x})\right),
\end{equation}
which makes use the property that the mean of the turbulent component is zero. 
In an effort to reduce the sensitivity of this method to potential measurement outliers, the bin-median is similarly computed. 
The bin-mean method also introduces error through gradients in the flow field over the range of blade positions over the bin width. 
One solution is to shrink the bin width, but the number of measurements $n$ occurring in the bin vary inversely with $\Delta \theta$. 
If $n$ is too small, the turbulent fluctuations have a non-zero mean. 
To reduce bin-width error while maintaining higher statistical certainty, a weighted average is computed, where the weight of each measurement varies inversely with its distance from the bin center:
\begin{equation}
\up(\bm{x},\theta_i) = \frac{\sum\limits_{j = 1}^n \left(\bm{u}(\bm{x},\theta_j)-\um(\bm{x})\right)\left|\theta_i+\Delta \theta - \theta_j\right|}{\sum\limits_{j = 1}^n \left|\theta_i+\Delta \theta - \theta_j\right|}.
\end{equation}
In these methods, each half rotation of the rotor is assumed to be one period of flow oscillation due to the symmetry of the two-bladed rotor. 
Reconstruction error, when compared with the full flow field, was minimized with $\Delta \theta = 3^\circ$, or 30 bins per half-revolution, resulting in, on average, $n = 67$ flow snapshots per bin.

As none of these approaches is entirely satisfactory in handling the error between the actual blade position and the position of the bin center introduced by free-running data acquisition, this motivates an exploration of alternative methods. 

\subsubsection{Fourier Series Reconstruction}

A Fourier-series based reconstruction of $M$ harmonics is given by
\begin{equation}
	\up(\bm{x},\theta(t)) = \sum\limits_{m = 1}^M A_m(\bm{x}) \sin\left[m \omega_b t + \phi_m(\bm{x})\right],
\end{equation}
wherein a series of sinusoidal functions are fit to the data. 
The resulting function is used to reconstruct the data at the bin-center blade positions, eliminating the inter-bin blade position error of the previous method.

In the case of this flow, selection of the base oscillation frequency, $\omega_b$, is simple given that the blade passing frequency is the primary driver of flow oscillations. 
The flow-field is computed, similar to the conditional average methods, by reconstruction at times $t_i$ that correspond to bin centers $\theta_i$. 
In practice, the coefficients and phase fields $A_m(\bm{x})$ and $\phi_m(\bm{x})$ are computed via the windowed fast Fourier transform, with the time series padded appropriately to ensure the FFT output includes all $r$ frequencies exactly equal to $m \omega_b$, removing potential frequency interpolation error. 
This is referred to in the following sections as the discrete Fourier transform (DFT) method. 
We evaluate two version of this method. 
First, for the ``DFTc'' method, the base oscillation frequency (the blade passing) is calculated from the average location of the largest peak of the flow data spectra. 
Second, in the ``DFTm'' method, the base frequency is calculated from the encoder data collected during turbine operation.

 \subsubsection{Multi-Modal Decomposition via Optimized DMD}
 In many cases, the base oscillation frequency may be unknown, or the flow may exhibit features that oscillate at unrelated frequencies. 
 In the case of a cross-flow turbine wake, the blade passing frequency may not be the only mechanism determining the time-scale of periodic fluctuations. 
 This motivates a generalization of the triple decomposition to Eq.~\eqref{eq:mm_triple} as introduced by~\cite{baj_2015}. 
 Here $\up$ is split into fluctuating components whose frequencies are not necessarily related, allowing this data-driven triple decomposition method to be used as an exploratory/diagnostic tool. 
 In this work, we used dynamic mode decomposition to identify the fluctuating components. 
 A related method that could be used is spectral POD, an implementation of the original POD of~\cite{lumley_1967}, with the "spectral POD" terminology introduced by~\cite{picard_2000}. A detailed account of the relationship between spectral POD and DMD is given by~\cite{towne_2018}.

The dynamic mode decomposition was introduced by~\cite{schmid2010jfm} in the fluids community to identify spatiotemporal coherent structures from time-series data. 
In its simplest form, the DMD algorithm extracts the dominant eigenvalues and eigenvectors of the best-fit linear operator that approximately advances the measured state forward in time.  
The DMD algorithm starts with two snapshot matrices constructed of spatial and temporal flow components:
\begin{align}
\begingroup
\setlength\arraycolsep{2pt}
\mathbf{X} = \begin{bmatrix} u(\bm{x}_1,t_1) & u(\bm{x}_1,t_2) & & u(\bm{x}_{1},t_{m-1}) \\ \vdots & \vdots & & \vdots \\ u(\bm{x}_n,t_1) & u(\bm{x}_n,t_2) & & u(\bm{x}_n,t_{m-1}) \\ v(\bm{x}_1,t_1) & v(\bm{x}_1,t_2) & & v(\bm{x}_{1},t_{m-1}) \\ \vdots & \vdots & {{\cdot}\mkern1mu{\cdot}\mkern1mu{\cdot}} & \vdots \\ v(\bm{x}_n,t_1) & v(\bm{x}_n,t_2) &  & v(\bm{x}_n,t_{m-1}) \\ w(\bm{x}_1,t_1) & w(\bm{x}_1,t_2) & & w(\bm{x}_{1},t_{m-1}) \\ \vdots & \vdots & & \vdots \\ w(\bm{x}_n,t_1) & w(\bm{x}_n,t_2) & & w(\bm{x}_n,t_{m-1}) \\
\end{bmatrix},
\mathbf{X}' = 
\begin{bmatrix} u(\bm{x}_1,t_2) & u(\bm{x}_1,t_3) & & u(\bm{x}_{1},t_m) \\ \vdots & \vdots & & \vdots \\ u(\bm{x}_n,t_2) & u(\bm{x}_n,t_3) & & u(\bm{x}_n,t_{m}) \\ v(\bm{x}_1,t_2) & v(\bm{x}_1,t_3) & & v(\bm{x}_{1},t_m) \\ \vdots & \vdots & {{\cdot}\mkern1mu{\cdot}\mkern1mu{\cdot}} & \vdots \\ v(\bm{x}_n,t_2) & v(\bm{x}_n,t_3) &  & v(\bm{x}_n,t_{m}) \\ w(\bm{x}_1,t_2) & w(\bm{x}_1,t_3) & & w(\bm{x}_{1},t_m) \\ \vdots & \vdots & & \vdots \\ w(\bm{x}_n,t_2) & w(\bm{x}_n,t_3) & & w(\bm{x}_n,t_{m}) \\
\end{bmatrix}.
\endgroup
\end{align}
The best-fit linear operator that maps $\mathbf{X}$ into $\mathbf{X}'$ is given by $\mathbf{A}$, satisfying the approximate relationship:
\begin{align}
\mathbf{X}'\approx \mathbf{A}\mathbf{X}.
\end{align}
In practice, this matrix $\mathbf{A}$ may be approximated using the pseudo-inverse of $\mathbf{X}$, which is computed by taking the singular value decomposition $\mathbf{X}=\mathbf{U}\boldsymbol{\Sigma}\mathbf{V}^T$ and inverting each of the matrices $\mathbf{U}$, $\boldsymbol{\Sigma}$, and $\mathbf{V}^T$:
\begin{align}
\mathbf{A} = \mathbf{X}'\mathbf{V}\boldsymbol{\Sigma}^{-1}\mathbf{U}^T.
\end{align}
The matrix $\boldsymbol{\Sigma}$ is diagonal, and both $\mathbf{U}$ and $\mathbf{V}$ are unitary, so their transposes are their inverses. 
However, if the state $\mathbf{X}$ is a large discretized fluid velocity or vorticity field, the matrix $\mathbf{A}$ may be intractably large to represent, let alone to analyze. 
Instead, we compute the projection of $\mathbf{A}$ onto the leading POD modes, given by the first $r$ columns of $\mathbf{U}$, denoted by $\mathbf{U}_r$:
\begin{align}
\tilde{\mathbf{A}} = \mathbf{U}_r^T\mathbf{A}\mathbf{U}_r = \mathbf{U}_r^T\mathbf{X}'\mathbf{V}_r\boldsymbol{\Sigma}_r^{-1}.
\end{align}
The matrices $\mathbf{A}$ and $\tilde{\mathbf{A}}$ share the same eigenvalues, so it is possible to compute the spectrum of $\mathbf{A}$ by computing the eigendecomposition of $\tilde{\mathbf{A}}$:
\begin{align}
\tilde{\mathbf{A}}\mathbf{W} = \mathbf{W}\boldsymbol{\Lambda}
\end{align}
where $\mathbf{W}$ contain the eigenvectors of $\tilde{\mathbf{A}}$ and $\boldsymbol{\Lambda}$ contains the eigenvalues.  
Finally, it is possible to compute the high-dimensional eigenvectors $\boldsymbol{\Phi}$ of the matrix $\mathbf{A}$ (e.g., the DMD modes), from the low-dimensional eigenvectors $\mathbf{W}$ using the \emph{exact} DMD algorithm of~\cite{tu2014jcd}:
\begin{align}
\boldsymbol{\Phi} = \mathbf{X}'\mathbf{V}_r\boldsymbol{\Sigma}_r^{-1}\mathbf{W}.
\end{align}
DMD has recently been connected to spectral POD~\citep{towne_2018}, used to analyze a cross-flow turbine wake by~\cite{araya_2017}, and the resolvent operator~\citep{sharma2016prf}. Another view is that DMD is an approximation of the Koopman operator, which is an infinite-dimensional linear operator that steps a system forward in time by operating on an infinite-dimensional Hilbert space of all scalar-valued functions of system measurements~\citep{rowley_2009, mezic_2013,kutz_2016}.

It is well known that the original DMD algorithm of~\cite{schmid2010jfm} is sensitive to noise~\citep{Bagheri2014pof} and there are several recent approaches to de-bias the algorithm for noisy data~\citep{dawson2016ef,hemati2017tcfd,askham2018siads}. 
The optimized DMD (optDMD) algorithm of~\cite{askham2018siads} considers the evolution of all of the snapshots at once, instead of through a single iteration through the map $\mathbf{A}$, and provides an efficient way of solving a nonlinear least-squares regression problem using variable projection. 
This has the added benefit of allowing for an optimal DMD fit from data that is unevenly spaced in time. 
This is, in general, a non-convex procedure, although there are efficient algorithms to compute this optimization, and the results indicate considerable noise robustness over standard algorithms.  

The optimized DMD method also provides a mechanism for constraining the eigenvalues of the returned modes, for example to keep them on the unit circle. 
This allows for solving of periodic-only optDMD modes, and can be used to restrict the oscillation frequencies. 
The data taken in these experiments consists of overlapping fields-of-view taken at separate times. 
When optDMD is performed on the entire dataset, the resulting modal oscillations are out-of-phase. 
The field-of-view overlap regions are used to correct the phase misalignment, resulting in full-field DMD modes. 
This method is likely useful for modal analysis in any experiment utilizing multiple overlapping measurement areas. 
Details of this method can be found in~\cite{Nair2020prf}. 

The ability to restrict DMD eigenvalues to lie on the unit circle is critical for application to the triple decomposition, where such behavior is inherent in the definition of the oscillatory term. 
The standard DMD algorithm could be used to determine oscillatory flow components by either selecting modes with imaginary-only eigenvalues, or by manually zeroing the real part of the eigenvalues. 
However, the original mode shapes returned by exact DMD are no longer guaranteed to best represent the data given the now altered eigenvalues. Optimal DMD circumvents this issue by iteratively optimizing the mode shapes given constraints on the eigenvalues. 

\subsection{Decomposition Methods Comparison}
 For each of the algorithms above, the periodic component is extracted, reconstructed for the full length of the original data set, and then added back to the mean flow ($\um + \up$). 
 The reconstruction is compared to the original flow in two ways. 
 First, the average $L_2$ error between the reconstruction and original data is computed. 
 A smaller $L_2$ error indicates that more of the oscillatory mode information is being captured. 
 However, as the original data contains turbulent fluctuations, this $L_2$ error is never identically zero. 
 Second, the total sum of the flow kinetic energy over space and time is computed for the original and reconstructed flow. 
 The ratio of these energies is indicated by the vertical axis of \fref{psd_err}\fl{a}, while the error is on the horizontal axis. 
 The DMD-based method results in a reconstructed flow field with more energy explained and a lower error. 
 Somewhat surprisingly, the DFTm method results in an order of magnitude higher error than DFTc as small errors in the encoder-measured frequency versus the true frequency results in large oscillation phase errors during reconstruction. 
 This illustrates the importance of knowing or calculating the base frequency of interest exactly when using a DFT-based method.
 Of the averaging methods, a bin median, which is resilient to outliers, explains more of the flow kinetic energy than a bin mean or weighted bin mean.

\begin{figure}
\centering
\includegraphics[width=\textwidth]{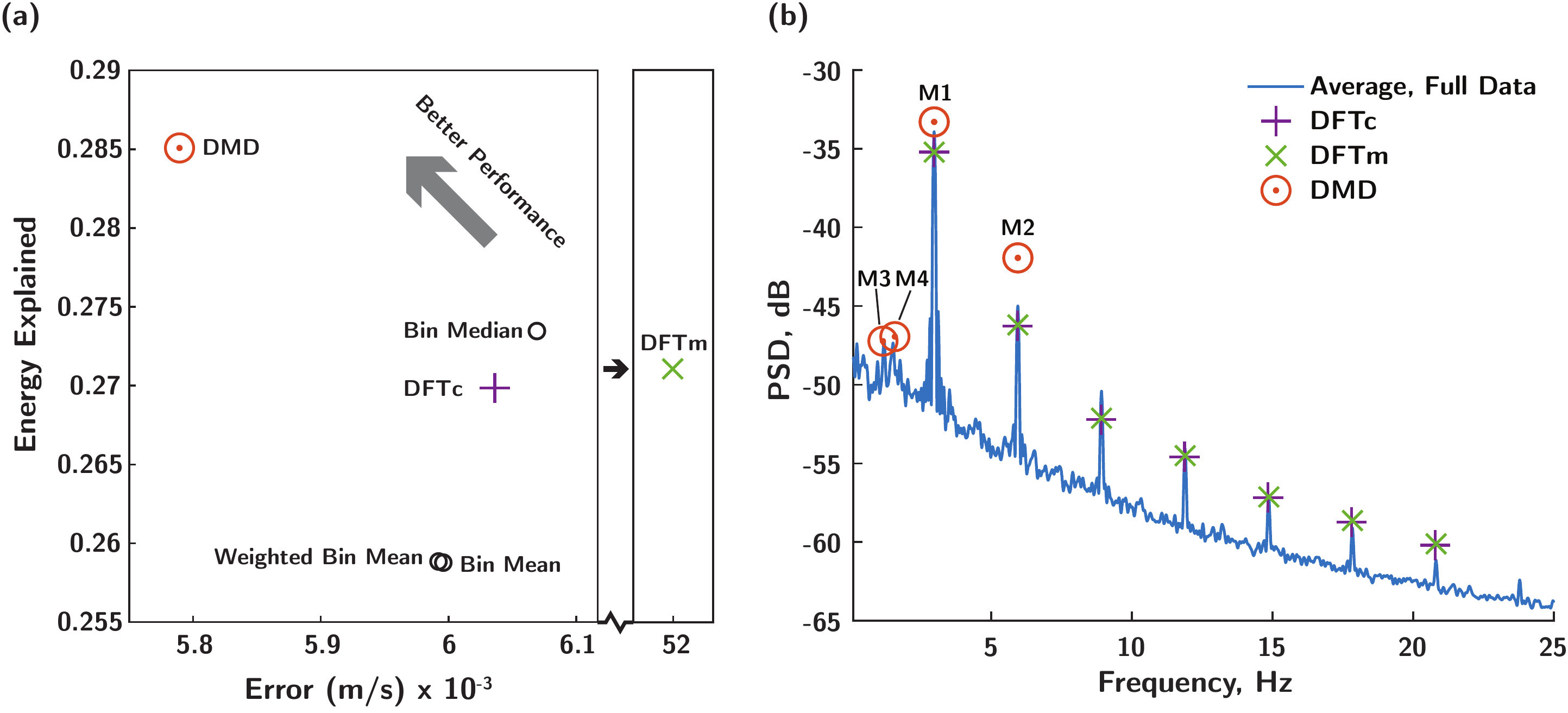}
\vspace{-.35in}
\caption{\fl{a} Kinetic energy content of the mean plus the reconstructed periodic flow normalized by the kinetic energy of the full flow measurements versus the $L2$ error of the reconstruction versus the original flow. We expect the most effective triple decomposition method to minimize the error while maximizing the amount of energy capture (as indicated by the arrow). \fl{b} Power spectra of the modes of the DFT and DMD methods. DMD indicates importance of low-frequency modes that may not be discovered by other methods. The first four DMD modes are labeled \textbf{\sffamily{M 1$\to$ 4}}.}\label{fig:psd_err}
\end{figure}

The DMD method does not require \textit{a priori} knowledge of the base frequency, and can be used to uncover flow phenomena oscillating at related or unrelated frequencies. 
As illustrated by \fref{psd_err}\fl{b}, the first seven modes extracted by the optDMD algorithm contain the blade passing frequency and its first harmonic, followed by five lower-frequency modes. 
To view these modes in terms of the full wake, not just the separate fields-of-view collected, the phase of oscillation of each mode in each individual field-of-view was adjusted via numerical optimization to match the oscillation of neighboring fields-of-view. 
This process is illustrated in \fref{phase_correct} and details are given in~\cite{Nair2020prf}.

\begin{figure}
\centering
\includegraphics[width=\textwidth]{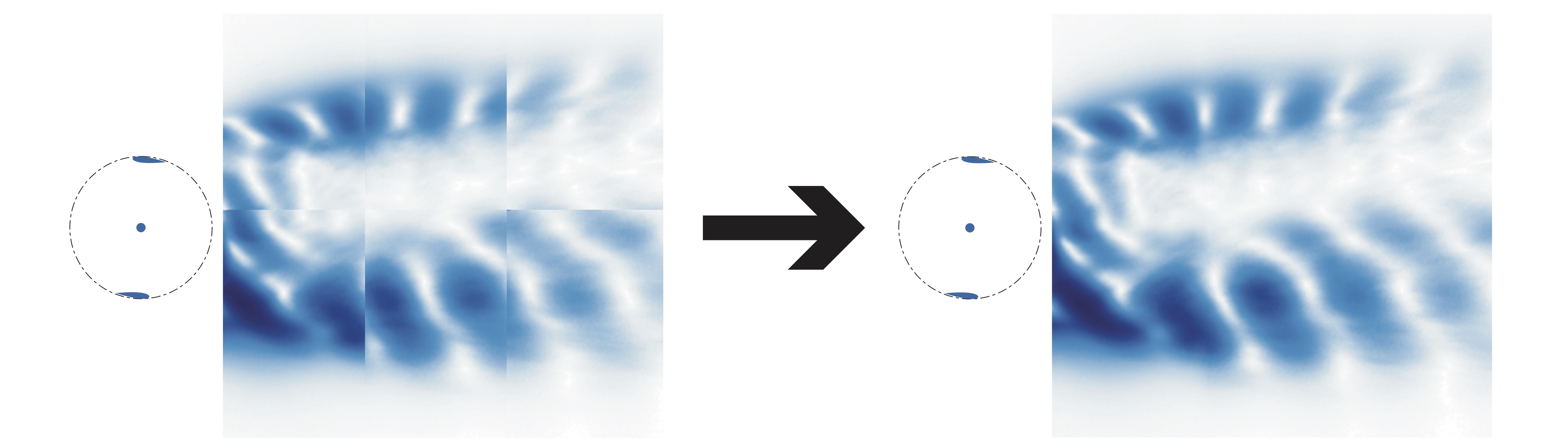}
\vspace{-.35in}
\caption{DMD mode phase correction process shown on the first DMD mode of the turbine wake data. Because data was collected at differing times, the phase of oscillation of the same mode in differing fields-of-view are not aligned. A numerical minimization of the error in field-of-view overlap regions is used to correct the phase. For this example, this is a five variable optimization problem (one field-of-view is the reference). }\label{fig:phase_correct}
\end{figure}

\subsection{Wake DMD Modes}
\begin{figure}
\centering
\includegraphics[width=\textwidth]{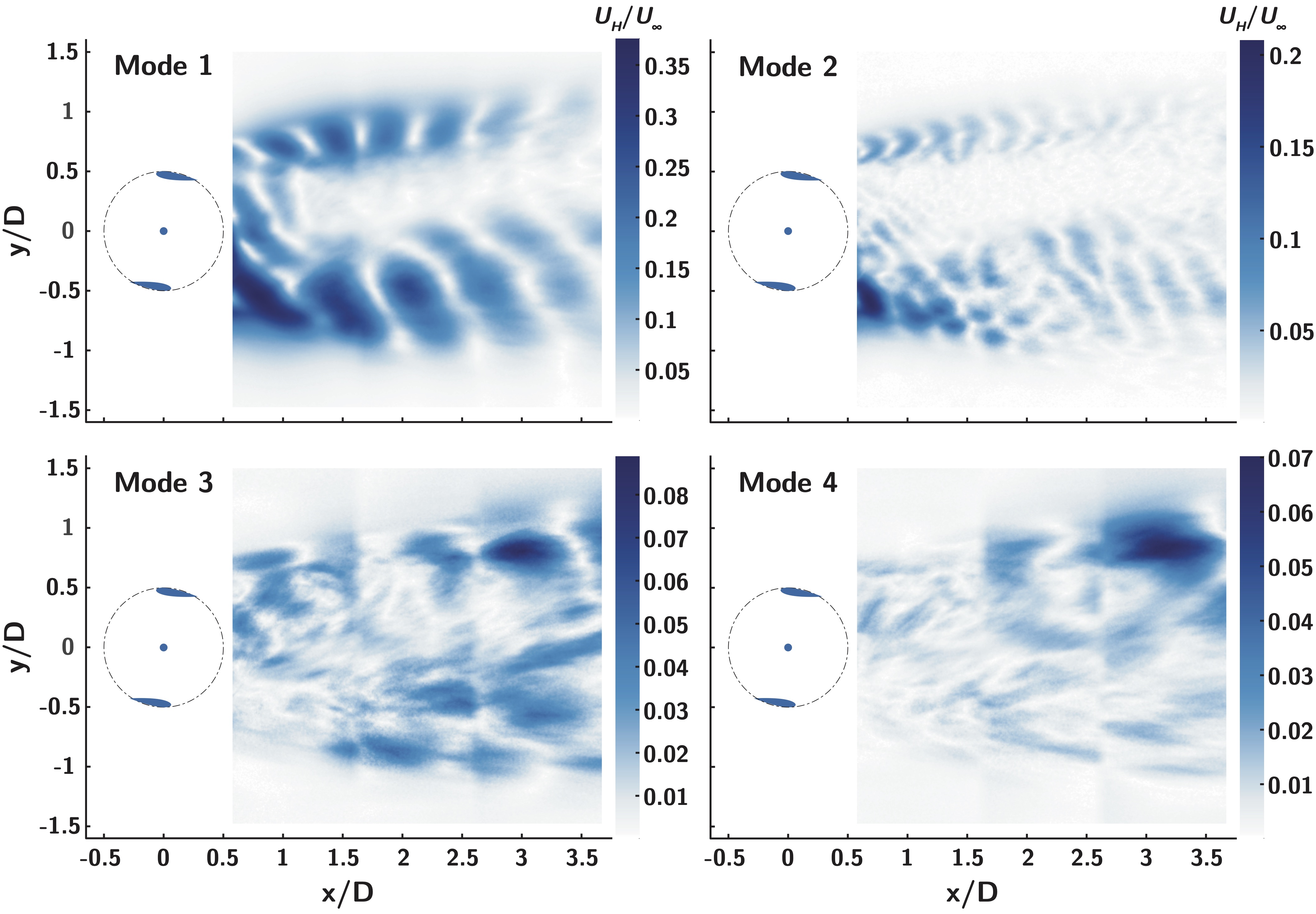}
\vspace{-.35in}
\caption{Modes extracted using the optDMD algorithm, with the phase of oscillations corrected. Modes are ranked by energy content and are identified by the \textbf{\sffamily{M}} labels in \fref{psd_err}\fl{b}.}\label{fig:dmd_modes}
\end{figure}
The optDMD triple decomposition extracts and ranks oscillatory modes in terms of energy content. The first four modes are shown in \fref{dmd_modes} and their corresponding frequencies are in \fref{psd_err}\fl{b}. Modes are computed using all three velocity components. The horizontal velocity magnitude
\begin{equation}
U_H = \sqrt{u^2 + v^2}
\end{equation}
is plotted in \fref{dmd_modes}. As mentioned previously, modes one and two correspond to the blade pass frequency and its first harmonic. 
The first harmonic energy content decreases faster in the downstream direction than the fundamental frequency, perhaps due to the more rapid dissipation of the smaller-scale features. 
The structures responsible for the energetic oscillations at the blade passing frequency will be discussed in the following section. 
The frequency of the third mode is half the blade passing frequency. 
Close to the turbine, energy in this mode illustrates changes in flow due to small geometric differences in the rotor blades or their mounting angle. 
Together, modes three and four illustrate a phenomenon on the advancing side of the wake. 
Structures that initially occur at the blade passing frequency seem to be breaking down or combining into lower-frequency structures in a repeatable manner. 
This could be evidence of a transition toward the bluff-body far wake oscillation documented by~\cite{araya_2017}. 
However, the frequency of mode four is 1.18 Hz, while the predicted bluff-body frequency for a cylinder of the same diameter as the rotor is 0.8 Hz. 
It is possible that full transition to the bluff-body frequency has not yet occurred, and that measurements made further downstream would show lower dominant frequencies. 

\section{Results and Analysis: Coherent Structures}
To provide a detailed description of the dynamics of coherent structures in the non-turbulent component of the wake ($\um + \up$), we present the Finite-time Lyapunov exponent (FTLE) fields~\citep{haller2002pof,shadden2005pd,green_2007,farazmand2012chaos}, in particular the Lagrangian coherent structure (LCS) ridges of the FTLE field. 
FTLE detects coherent structures via their boundaries by computing the maximum strain rate of a Lagrangian packet of fluid over a finite time period. 
Integration backward and forward in time yields attracting and repelling FTLE ridges, respectively, which enclose coherent structures. 
Unlike Eulerian methods, such as the Q-Criterion~\citep{hunt_1988}, the FTLE field does not require a user-defined, subjective threshold to identify coherent structures. 
Additionally, as an integration-based method, FTLE analysis is more robust to noise than derivative-based methods~\citep{green_2007}.  
Though other methods have been shown to identify LCS more rigorously~\citep{haller_2013}, the ridges of the FTLE field remain practically useful due to their ease of computation, mathematical simplicity, and conceptual accessibility. 
Here, structures associated with shear-layer roll-up, blade-vortex shedding, and bluff-body wake oscillation are identified, and the role of these structures in wake mixing and recovery is considered. 
In addition, vortex-core tracking is performed on the full time-resolved wake to determine vortex longevity and trajectory repeatability, which has implications for interactions with downstream turbines in an array.

\subsection{Computing the Finite-Time Lyapunov Exponent}

Lagrangian coherent structures are useful for identifying coherent regions of unsteady fluid flows that are segmented by time-varying separatrices, which are the unsteady analogues of stable and unstable invariant manifolds in dynamical systems~\citep{haller2002pof,shadden2005pd}. 
LCS are often computed as the second derivative ridges of the FTLE field~\citep{shadden2005pd}, which describes the maximum rate of stretching of a Lagrangian packet of fluid over a finite time period.  
The more recent method of~\cite{farazmand2012chaos} uses variational theory to compute LCS from FTLE fields.

The FTLE field is generally computed by integrating passive tracer particles along the flow of the velocity field $\bm{u}(\bm{x},t)$ for a time span $T$ as:
\begin{align}
\Phi_{t_0}^T\left(\bm{x}(t_0)\right) = \bm{x}(t_0+T) = \bm{x}(t_0) + \int_{t_0}^{t_0 + T} \bm{u}(\bm{x}(\tau),\tau)\,d\tau,
\end{align}
where $\Phi_{t_0}^T$ is the flow map.  
Next, the flow map Jacobian, $D\Phi_{t_0}^T$ is approximated via finite-difference derivatives with neighboring points in the flow.  
In two-dimensions, the flow map Jacobian at a point $\mathbf{x}_{i,j}$ is:
\begin{align}
\renewcommand\arraystretch{2}
\left(D\Phi_{t_0}^T\right)_{i,j} \approx \begin{bmatrix} 
\frac{x(t_0 + T)_{i+1,j}-x(t_0 + T)_{i-1,j}}{x(t_0)_{i+1,j}-x(t_0)_{i-1,j}} 
& \frac{x(t_0 + T)_{i,j+1}-x(t_0 + T)_{i,j-1}}{y(t_0)_{i,j+1}-y(t_0)_{i,j-1}}\\
\frac{y(t_0 + T)_{i+1,j}-y(t_0 + T)_{i-1,j}}{x(t_0)_{i+1,j}-x(t_0)_{i-1,j}} 
& \frac{y(t_0 + T)_{i,j+1}-y(t_0 + T)_{i,j-1}}{y(t_0)_{i,j+1}-y(t_0)_{i,j-1}}
\end{bmatrix}.
\end{align}
From a continuum mechanics standpoint, this is a numerical computation of the deformation gradient. 
The finite-time Lyapunov exponent $\sigma$ is computed from the largest eigenvalue $\lambda_\text{max}$ of the Cauchy-Green deformation tensor $\bm{\Delta}=\left(D\Phi_{t_0}^T\right)^\intercal D\Phi_{t_0}^T$, 
which is the maximum singular value of the flow map Jacobian:
\begin{align}
\sigma(\bm{x}_0,t_0,T) = \frac{1}{T}\ln\left(\sqrt{\lambda_\text{max}\left[\bm{\Delta}(\bm{x}_0,t_0,T)\right]}\right).
\end{align}
The variable $\sigma$ is a scalar field that is typically computed on a discrete grid of particles, and for unsteady flows this field is recomputed at every time step, introducing redundant computations that may be eliminated~\citep{brunton2010chaos,luchtenburg2014jcp}.  
When $\sigma$ is large, then neighboring particles undergo considerable stretching along the flow, while particles with small $\sigma$ will tend to remain in coherent patches with their neighbors.  
Repelling or attracting FTLE field structures may be computed by integrating particles either forward or backward time, respectively.  

As the integration period, $T$, is increased, the FTLE ridges become more defined, though their locations remain constant~\citep{green_2007}. 
For studies with limited interrogation windows, such as this one, this can present a problem. 
Increasing integration time results in more finely resolved FTLE ridges, but increases the chance that a passive tracer will exit the flow-field before the end of the integration. 
A partial solution to this issue is to subtract the global (time and space) mean velocity vector from the flow field, as this will not affect the mechanics of a Lagrangian fluid packet. 
However, with flows that exhibit local velocities differing significantly from the global mean, this strategy is only partially effective since the local velocities can still eject the passive tracers. 
FTLE computations presented here assign passive tracers that leave the domain their FTLE value at the time of exit, resulting in a decrease in FTLE ridge definition near some domain edges. 

\subsection{Wake Coherent Structures}
In addition to analyzing the frequencies of different components of the oscillating wake, it is illuminating to examine the time evolution of periodic coherent structures. 
In \fref{ftle} we show the forward- and backward-time FTLE fields computed from the mean and the reconstructed periodic component (optDMD method). 
We observe the structures driving the flow oscillation that were evident in the first two DMD modes. 
On the advancing side of the wake ($y/D > 0$), there is a vortex street with vorticity opposite the direction of turbine rotation (\fref{lcs_vort}). 
Roll up of the strong shear layer in the flow, as shown in \fref{mean_wake}\fl{a} and \fl{b}, is likely the energetic source for these vortices, while the disturbance caused by the blade passage ensures the roll-up occurs at a regular frequency. 
It is possible that the blade sheds some vorticity while advancing as well, but the arrangement of the shear layer means this is not necessary for vortex roll-up.
\begin{figure}
\centering
\includegraphics[width=\textwidth]{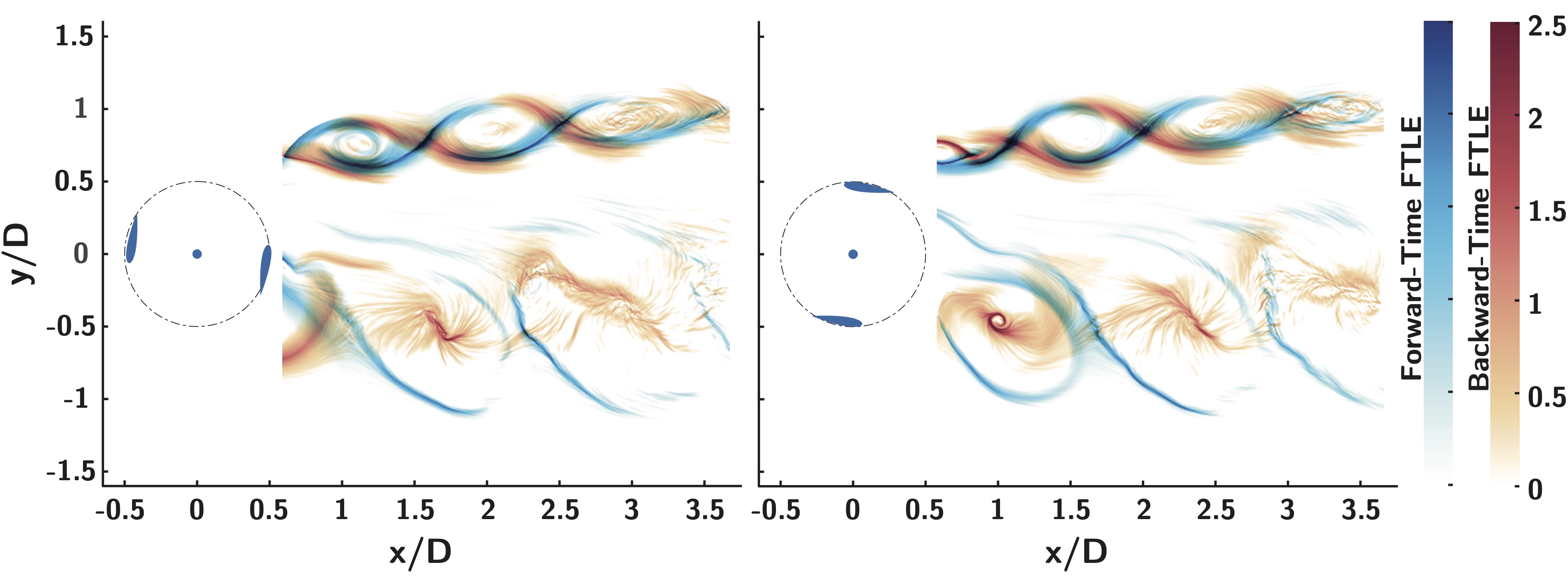}
\vspace{-.35in}
\caption{Forward and backward FTLE fields computed on $\um + \up$ (optDMD method). These fields represent areas of maximum stretch and convergence respectively, and together outline the boundaries of coherent structures.}\label{fig:ftle}
\end{figure}

\begin{figure}
\centering
\includegraphics[width=\textwidth]{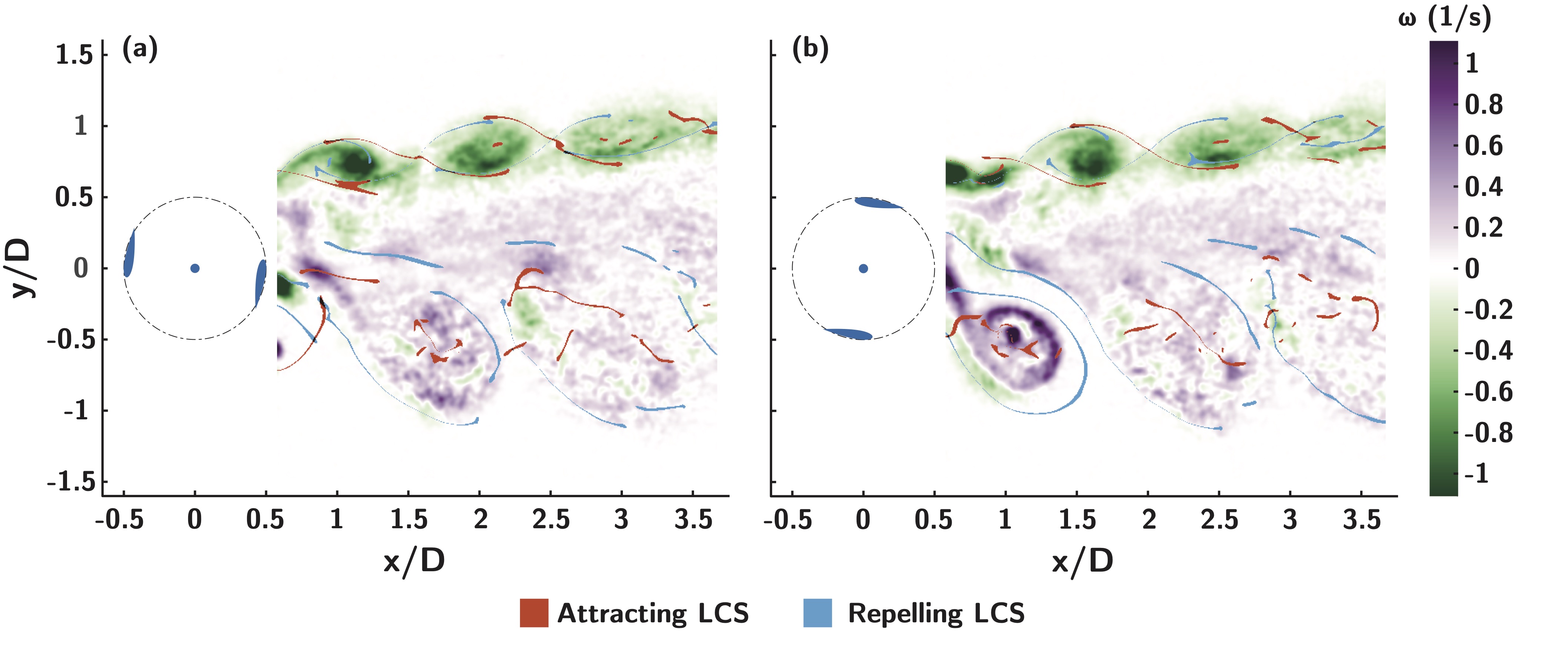}
\vspace{-.35in}
\caption{Second derivative ridges of the FTLE field superimposed on the out-of-plane flow vorticity (the curl of the horizontal velocity components).}\label{fig:lcs_vort}
\end{figure}

On the retreating side of the wake ($y/D < 0$), a large vortex structure is also apparent, but it is due to a completely different mechanism. 
During the power-producing portion of the blade stroke (centered around $\theta = 90^\circ$ when when the blade is farthest upstream) lift production requires the generation of counter-clockwise circulation around the blade. 
Because this circulation is not permanent, but is created every rotation, packets of opposite circulation must be shed into the wake~\citep{battisti_2011}. 
Detailed analysis of the actual formation of this vortex structure on the blade would require further measurements upstream, but the vortex has the correct sign to be a trailing edge vortex perhaps partly in response to a leading edge vortex formed during dynamic stall. 
This structure dissipates more quickly than the vortex street on the advancing side, perhaps in part due to the lack of a strong energizing shear layer on this side of the wake. 
At the same time, horizontal (in-plane) mixing caused by this structure may contribute to the lesser shear layer and faster wake recovery on the retreating side (see \fref{mean_wake}\fl{a}). 

\begin{figure}
\centering
\includegraphics[width=\textwidth]{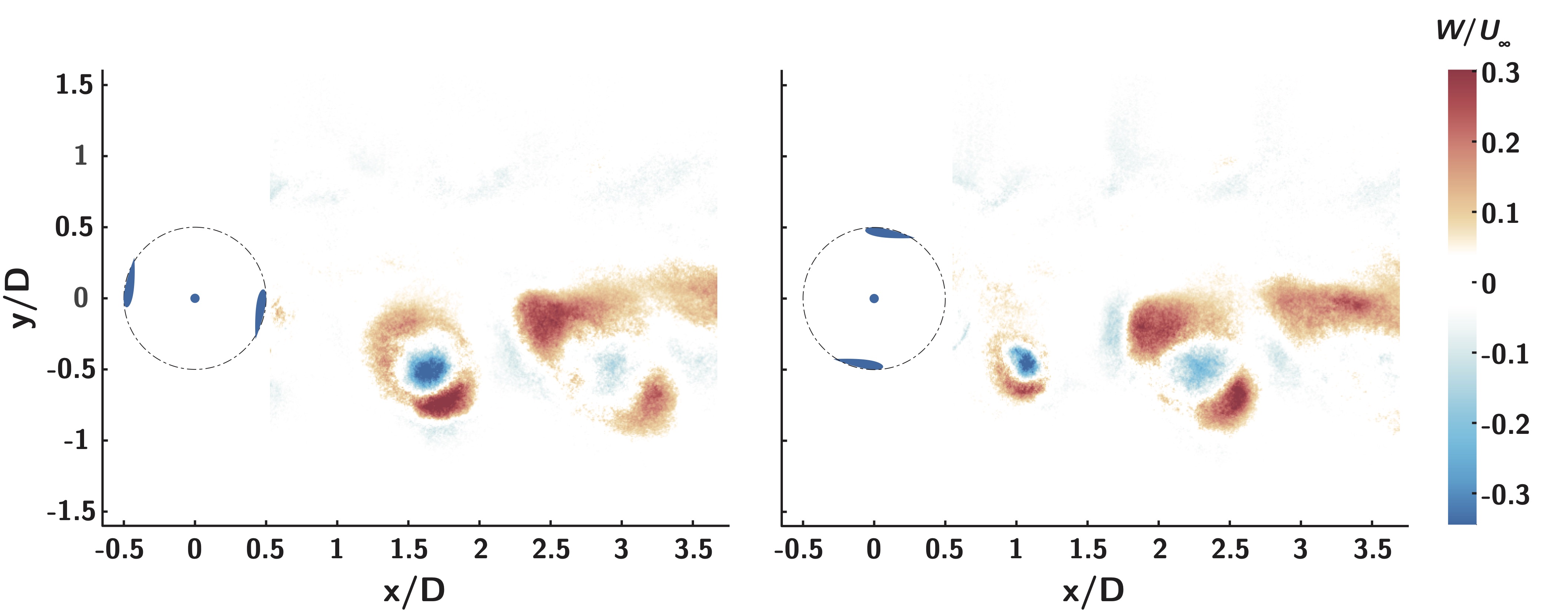}
\vspace{-.35in}
\caption{Out-of plane (vertical) velocity, mean and periodic component ($\bar{w} + \tilde{w}$).}\label{fig:wake_w}
\end{figure}

The vortex structure shed on the retreating side is more complex than a single rotating packet of fluid. 
An intense core of vorticity is surrounded by a ring of vorticity of the same rotation direction. 
This unstable arrangement may be partially responsible for the rapid breakdown of this structure, though to see why it survives at all we examine the out-of-plane velocity ($w$)  in \fref{wake_w}. 
The inner vortex core has intense axial (vertical) velocity in the negative direction, while the outer vorticity ring has vertical velocity towards the flume free surface. 
The out-of-plane velocity observed in the mean wake (\fref{mean_wake}\fl{d}) is entirely due to this structure. 
It appears that this structure already contains significant axial flow when it is shed into the wake, so it seems unlikely that it is due purely to asymmetries in the freestream and their influence on shed tip vortices. 
Axial flow in dynamic-stall related structures has been reported in insect flight~\citep{birch2001nature} and delta wing aircraft~\citep{wu_1991}. 
However, the source of the pressure gradient that drives vertical flow within a rotor that is symmetric about the mid-plane remains unknown. 

\begin{figure}
\vspace{-.1in}
\centering
\includegraphics[width=0.85\textwidth]{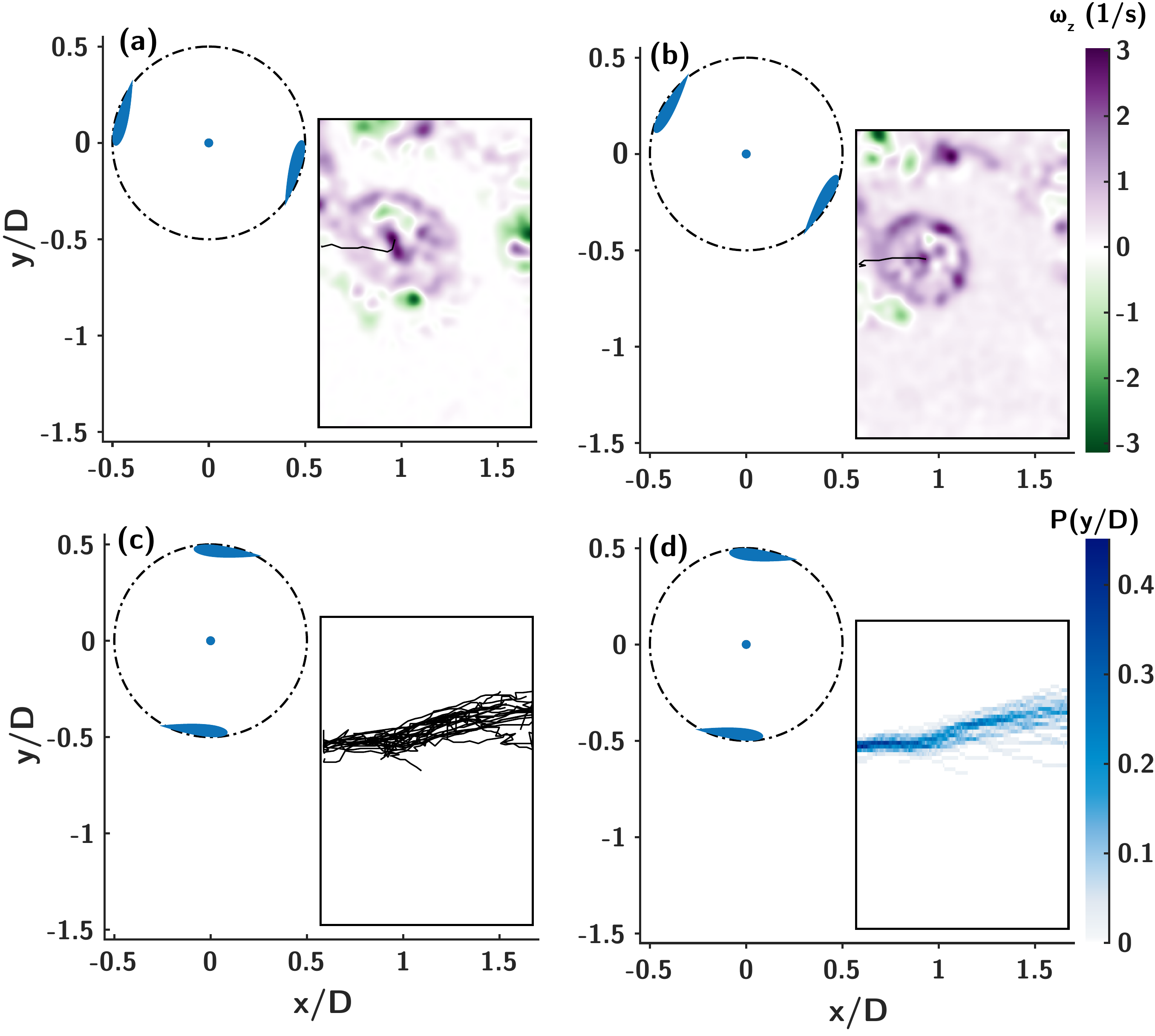}
\vspace{-.25in}
\caption{Retreating side vortex core tracking. \fl{a} and \fl{b}, example tracks and vorticity fields. \fl{c} All 50 tracks for the sampling period. \fl{d} Probability distribution of track $y/D$ location. }\label{fig:tracks}
\end{figure}

The predictability of the trajectories of wake coherent structures is of interest for optimizing the performance of closely spaced downstream turbines in an array. 
Consistent trajectories may make it easier for a downstream turbine to harness or avoid coherent structures (i.e., operation can be coordinated based on knowledge of upstream turbine phase). 
In \fref{tracks}\fl{a} and \fl{b}, the core of the retreating side vortex is tracked. 
All tracks are shown in \fref{tracks}\fl{c}, and in \fl{d}, the probability density of the cross-stream location of the vortex core is plotted as a function of streamwise vortex core position. 
The position of the vortex core increases in uncertainty with $x/D$. 
This indicates that the longevity of vortices may be greater than that predicted by just the periodic component of the flow, on which both the optDMD modes and, therefore, the subsequent FTLE are based. 
However, regardless of longevity, the rapid increase in the uncertainty of trajectories means that the structure could not be intercepted reliability by a downstream turbine blade for $x/D >1$. 

\section{Discussion and Conclusions}

Detailed analysis of a cross-flow turbine wake through stereo PIV measurements of the rotor mid-plane has been presented. 
First, we show that the mean wake structure is similar to prior investigations, despite the relatively higher $c/R$ for this turbine. 
Using phase-resolved performance measurements, we hypothesize the observed wake skew is simply a consequence of momentum conservation for the torque-producing tangential force acting on the blade, clarifying the conflicting and inconsistent explanations in prior work. 
Second, we show that the optDMD algorithm can identify the periodic flow in a triple decomposition with with lower error than conditional averaging or DFT-based methods, explains more of the energy present, and discovers oscillating structures at unrelated frequencies. 
As for~\cite{araya_2017}, we observe indications of a transition to bluff body shedding in the far wake, particularly on the advancing side of the rotor. 
Third, building on the identified periodic structures, we present the first detailed description of Lagrangian coherent structures in a cross-flow turbine wake. 
Vortex streets on the advancing and retreating sides are observed and their formation mechanisms hypothesized. 
We observe remarkably high axial flow in the core of the vortices associated with lift production on the retreating side of the wake and note that this axial flow appears to originate within the confines of the rotor. 
The presence of a mean vertical flow through the mid-plane of a span-wise symmetric rotor motivates future investigation of its origins and connecting the near-wake structures observed here to the fluid-structure interactions at the blade.

As described in \textsection \ref{sec:intro}, one of the factors driving modern interest in cross-flow turbine fluid dynamics is the potential for superior performance in arrays, particularly arrays of closely-spaced turbines where interaction with mean and periodic coherent structures in one turbine's wake can be exploited by another. 
As shown here, the mean and periodic wake structure is temporally and spatially rich, with varying implications for turbine arrays. 
For example, coherent structures shed on the retreating side of the wake likely dissipate too quickly to interact with a turbine blade more than one diameter downstream. 
The coherent structures on the advancing side wake are more persistent and transition to lower frequency structures. 
As these structures are similar to the rotor size and located in the slowest mean flow, they would likely be detrimental to downstream turbine performance at greater distances. 
We do note, however, while the coherent structures are visually striking (see \fref{ftle}), the energy contained in these structures is quite low (c.f., magnitude of DMD mode 4 in \fref{dmd_modes}). 
Finally, we note that this dataset is at a Reynolds number that is lower than in many commercial applications such that an increase in Reynolds number may increase the dissipation rate of coherent structures, as well as decrease consistency in their trajectory and strength~\citep{rocha_2018}.

\section*{Acknowledgements}
BS' current affiliation is XFlow Energy Company, Seattle, WA 98108. BS and BP would like to acknowledge support from the U.S. Department of Defense Naval Facilities Engineering Command under N00024-10-D-6318 (Task Order 0037) as well as funding from the U.S. Department of Energy (DE-EE006816) and Washington Clean Energy Fund (15-92201-006) for the PIV system acquisition and the Alice C. Tyler Charitable Trust for upgrades to the experimental facility. We would like to thank Craig Hill for his role in PIV system commissioning and data acquisition. SLB would like to acknowledge support from the Army Research Office (ARO W911NF-19-1-0045; Program Manager Matthew Munson).  

\begin{spacing}{.95}
\setlength{\bibsep}{2.6pt}

\end{spacing}
\end{document}